# Self-driven water or polarized liquid based ultraviolet photodetector

*Chang Liu[1], Yanghua Lu[1,2], Xutao Yu[1], Can Wang[1], Runjiang Shen[1] and Shisheng Lin[1,2,3]\**

[1]College of Information Science and Electronic Engineering, Zhejiang University, Hangzhou, 310027, P. R. China.

[2]Hangzhou Gelanfeng Technology Co. Ltd, Hangzhou, 310051, P. R. China.

[3]State Key Laboratory of Modern Optical Instrumentation, Zhejiang University, Hangzhou, 310027, P. R. China.

Email: shishenglin@zju.edu.cn

\*Corresponding author.

**Abstract**

Traditionally, photodetector is based on solid materials constructed PN junction, which needs many delicate growth technologies. Herein, we demonstrate the feasibility of polarized liquid triggered photodetector where the liquid is sandwiched between P-type or N-type semiconductor which can be chosen freely according to the requirement of the specific response wavelength. Under the multiple cycles of optical switching, transient photo-polarized current, steady state photo-polarized current and depolarized current are repeatably observed in semiconductor/polar liquids/semiconductor structure. The responsivity and specific detectivity of transient photo-polarized current in N-GaN/water/P-GaN reach values of 104.2 mA/W and $4.4\times10^{12}$ Jones at 365 nm, and 52.8 mA/W and $1.9\times10^{12}$ Jones at 254 nm illumination under zero voltage bias. We anticipate that our research will have a profound impact on integrating self-powered photodetector with freely selectable wavelength bands.

**Introduction**

Due to the Fermi level difference of P-type and N-type materials, the built-in field can block the otherwise diffused carriers, which achieves a balanced state named PN junction[1, 2, 3]. The working mechanism of Schottky diode is similar to PN diode by substituting one part of semiconductor material by metal. Those PN junctions and Schottky diode can function as photodetectors, where the incoming light can trigger extra electron-hole pairs separated by the built field subsequently, contributing the response of the electrical signal to the light[4, 5, 6, 7, 8]. As an indispensable part of photo detection, ultraviolet photodetector can be widely used in many application scenarios such as aerospace communication and outer space exploration[9, 10]. To date, many wide-bandgap semiconductor materials, such as $Ga_2O_3$[11], GaN[12, 13, 14, 15] have been investigated for semiconductor PN junction photodetector. However, the fancy modern growth technologies, such as metal organic chemical vapor deposition (MOCVD), molecular beam epitaxy (MBE), and magnetron sputtering techniques must be used to fabricate a good PN junction photodetector with high detectivity or responsivity. Besides, the top layer of grown material should have a lattice parameter close to the bottom layer, which limits the choice of semiconductors for a required response to different light sources.

As a novel generation of lightweight and high-efficiency in-situ energy acquisition technology, mechanical energy the moving water can be converted into electricity, which can be applied to the field of Internet of Things (IOT)[16, 17, 18]. In our previous work[19, 20], water molecules have been demonstrated to harvest energy as they can be polarized while moving in-between the P and N type semiconductors in dynamic diodes. However, the effect of photo-carriers in those dynamic diodes have not been investigated although the energy of photons can effectively couple with electrons in semiconductor and subsequently with the water molecules.

Following the above-mentioned physics, we discovered that inserting polar liquid into the PN junction can generate persistent photo-polarized current under illumination and a high-performance ultraviolet photodetector can be achieved. Besides, we can simply acquire the photodetector with response wavelength chosen flexibly as there is no limitation of the lattice match. Under the difference between the polar liquid chemical potential and the semiconductor Fermi level, the photogenerated electrons and holes will continuously move to the opposite sides of the polar liquid droplet, thus the photodetector will output stable photo-polarization current during the liquid polarization/depolarization process. Under zero bias voltage, the responsivity (R) and specific detectivity ($D^*$) of transient photo-polarized current in N-GaN/water/P-GaN are 52.8 mA/W and $1.9\times10^{12}$ Jones at 254 nm and 104.2 mA/W and $4.4\times10^{12}$ Jones at 365 nm, respectively. Our method reveals the polarizing process of polar liquids under the photoelectric effect, providing a novel and promising approach for converting input light energy into sustained polarized electricity, which is a promising way to integrate and miniaturize mix-phased UV optoelectronics.

## Results and Discussion

Figure 1a shows a 3D schematic diagram of designed device, where polar liquid is encapsulated by insulating tape, and polar liquid surfaces could contact the semiconductors. Electrode metals for ohmic contacts are Ni/Au for P-GaN and N-GaN. The structural properties of related semiconductors are confirmed by Raman spectroscopy. As shown in Figure 1b, the obvious characteristic positioned peaks at 569.53 and 735.33 cm$^{-1}$ are corresponding to $E_2$ (high) and $A_1$ (LO) phonon modes of GaN and peak located at 418.03 cm$^{-1}$ is corresponding to $Al_2O_3$ substrate[12, 21]. Meanwhile, Figure S1 (Supporting Information) shows the scanning electron microscope (SEM) image of fabricated GaN, uniform plane distribution in space could be observed clearly from the image. The I-V characteristics of representative N-GaN/W/P-GaN device was measured in the dark and under 365 nm irradiation, which exhibits remarkable rectification characteristics under dark state. It clearly shows the photo-current is significantly increased for three orders of magnitudes compared with dark current, where the dark current is $2.8 \times 10^{-4}$ μA and the photo-current is -0.484 μA at zero bias.

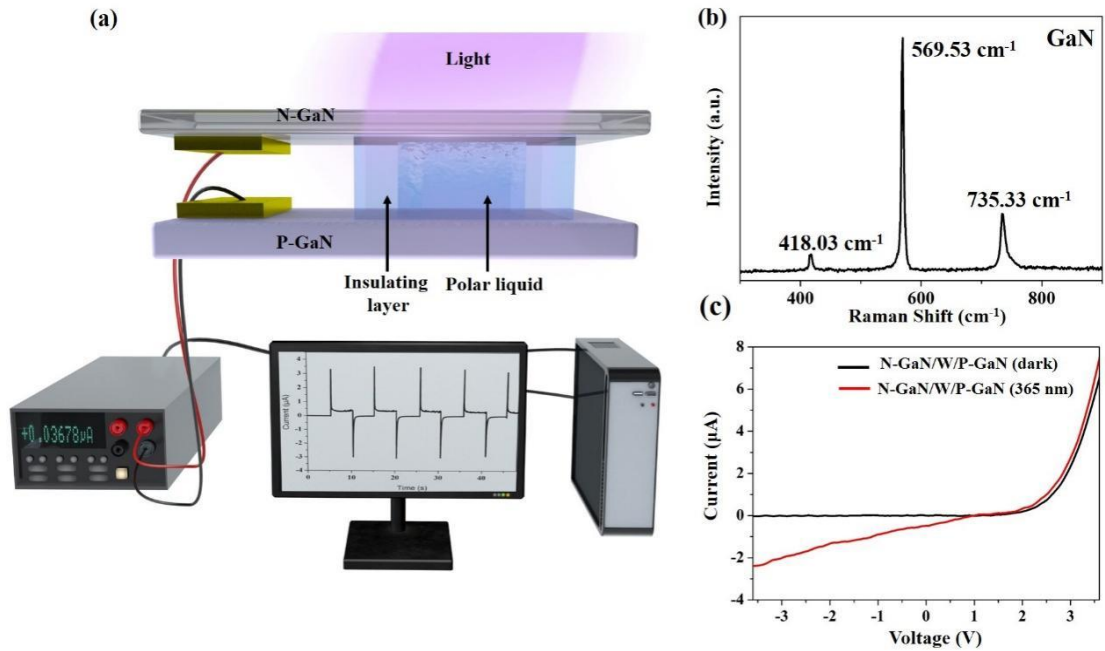

Figure 1. (a) Schematic diagrams of device structure and measurement system. (b) Raman spectrum of GaN. (c) I-V curves of the N-GaN/W/P-GaN device under dark environment and illumination of 365 nm light.

The working mechanism of the N-GaN/liquid/P-GaN photodetector under optical switching is illustrated in Figure 2. At the initial contact (Figure 2b), under the action of Fermi level difference in two semiconductors and chemical potential of water molecules, polar liquid molecules are instantly polarized. The water molecules will be polarized and arranged in an orderly manner, where oxygen atom and hydrogen atom end points to P-GaN and N-GaN, respectively. Our previous work of graphene-water-semiconductor direct current electricity generator also revealed the polarized process[19]. The majority carriers as electrons in N-type semiconductors will accumulate at the N-GaN/water interface, and the majority carriers of holes in P-type semiconductors will accumulate at the interface of P-GaN/water. Finally, the polarized molecule layer will screen the effect of electric field of PN junction and reaches electrostatic equilibrium with the Fermi level alignment[20].

Under illumination, the semiconductor will be instant excited by irradiation, and numerous carriers will gather at the interface between semiconductor and polarized liquid, leading to the polarization of some water molecules as shown in Figure 2c. At this time, the photogenerated electrons and holes will move to the interface between semiconductor and liquid under the effect of built-in field, and the movement of the carriers will generate the transient photo-polarization current, as indicated by the process of step 1 in Figure 2a. In the second step, while there are more and more water molecules are polarized, a dynamic polarization and de-polarization process will be stabilized and the steady photo-polarization current appears as shown as step 2 in Figure 2a. As the number of photo-generated carriers reach a maximum, the polarization current tends to be stable under continuous lighting, which is due to the equilibrium of the polarization and depolarization process of the water molecules under the excitation of thermal energy, leading to the continuous current output (Figure 2d). As indicated by step 3 of Figure 2a, a negative polarization current is generated when the light source is turned off and the physical picture is changed from Figure 2d to Figure 2e, which shows that most part of the water molecules are depolarized as a result of disappearing photo-induced carriers, leading to the negative current output. Primitive photo-excited carriers will instantaneously relax and recombine within the semiconductor, and the water molecules in the droplet revert to a disordered state, except for the two-phase interface (Figure 2d). In the fourth step, after the switching off for tens of milliseconds, the current almost recover to zero under the recombination of photo-excited carriers, the majority carriers in the semiconductor return to initial contact state.

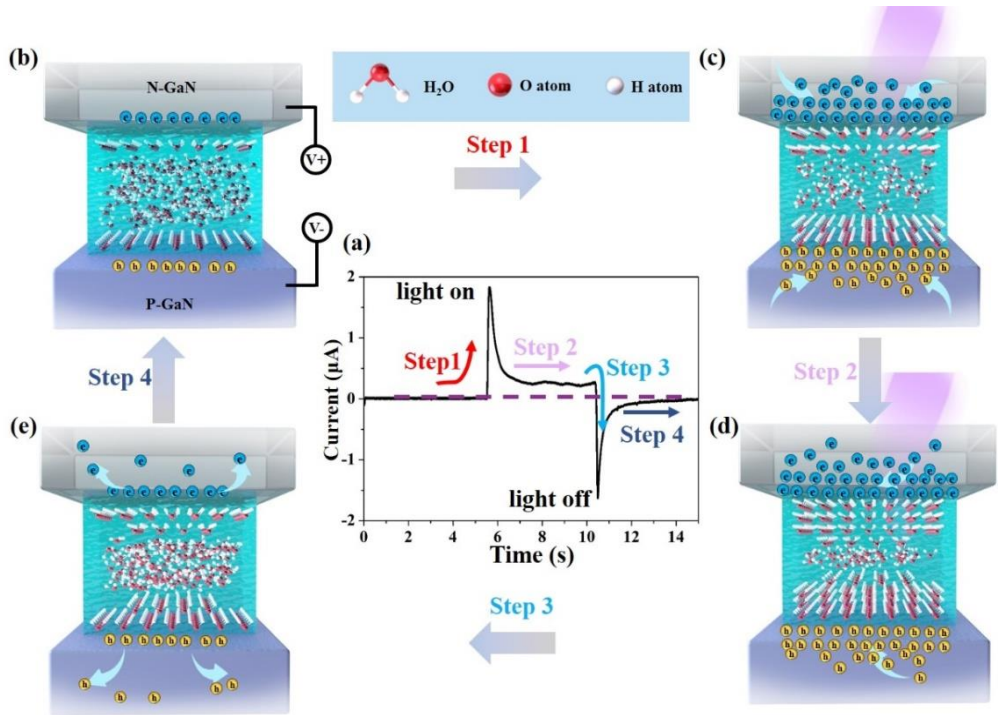

Figure 2. Working mechanism of the N-GaN/W/P-GaN photodetector under light illumination. (a) Enlarged view of time-current curves of device. (b) Equilibrium steady state of N-GaN/W/GaN photodetector after the contact. (c) Photocarrier accumulation induced the polarization state after the semiconductor excited. (d) Continuously stable polarization state induced by accumulation under stable illumination. (e) Depolarization process after turning off the light.

To present a comprehensive understanding of wavelength and liquid dependence, we performed current and time (I-T) curve tests under 0 V with multiple cycles of optical switching. I-T curves of N-GaN/liquid/P-GaN shows reproducible switching behavior in Figure 3a and Figure 3b, which reveals distinct polarization and depolarization processes. The transient/steady state photo-polarization current of N-GaN/W/P-GaN basically reached 1.86 μA as well as 0.232 μA under 254 nm irradiation, and 1.86 μA as well as 0.281 μA under 365 nm irradiation. Meanwhile, we also explored the I-T curves of the device adopting ethyl alcohol (EA) as the polar liquid, where a similar I-T curve can be produced compared with that adopting water as the polar liquid. The transient/steady state photo-polarization current of N-GaN/EA/P-GaN basically reached 2.79

µA/0.263 µA under 254 nm irradiation, and 3.38 µA/0.354 µA under 365 nm irradiation. Under the same light irradiation, the steady state current values of the N-GaN/EA/P-GaN photodetectors are slightly larger than N-GaN/W/P-GaN, as that EA has a smaller dielectric constant compared with water and thus more EA molecules can be polarized. Moreover, the transient and steady state photo-polarization current under the 365 nm irradiation become much larger than that under 254 nm. In case of GaN, the absorption band edge is around 365 nm (Figure S2 (Supporting Information)), so photons of suitable energy at 365 nm can excite more charge carriers in GaN compared with that of 254 nm, leading to more carriers transferred to the solid-liquid interface.

The polarization and depolarization process of the liquid is the key for the function of the mix-phased photodetector. To further prove that the photo-current is caused by polarization, different N-GaN/liquid/P-GaN systems are tested as the results shown in Figure 3c. As expected, the polar liquid including EA and water produced 3.36 and 1.86 µA photo-polarization current under 365 nm illumination. However, the nonpolar liquid n-hexane did not produce photo-polarization current under the same test conditions (enlarged output current curves was given in Figure S3 (Supporting Information)), confirming that the photocurrent is caused by polarization by the liquid phase in-between the semiconductors. It is worth noting that under the same wavelength of light irradiation, the transient photo-polarization current of ethyl alcohol is significantly larger than that of water. The relative permittivity $\varepsilon_r$ of ethyl alcohol and water is 37 and 78.3 F/m, respectively[20]. The material with a smaller dielectric constant has a weaker charge screening effect, that is, weaker the bound charge ability[22]. Therefore, the electric charges are more easily polarized in the same external electric field. This physical picture could be used to identify the relative magnitude of the dielectric constants of polar liquids.

In order to further understand the mechanism of carrier transport. The energy band structure of N-GaN/liquid/P-GaN device is shown in Figure 3d, where the electron affinity of GaN is around ~3.6 eV, and the work function of N-GaN and P-GaN is around ~4.0 eV and ~6.1 eV according to the doping concentration of the Si and Mg[23, 24, 25]. In case of water, the chemical potential ($\mu_e$) is 4.26 eV under standard atmospheric[26], which is located between the work function of N-GaN and P-GaN. As shown in the Figure 3e, the band gap alignment will occur between the interface after contact, and the electrons in N-GaN tendency diffuse to P-GaN and eventually accumulate at the solid-liquid surface of N-GaN/W. Similarly, the holes in P-GaN tendency to diffuse to N-GaN and eventually accumulate on the solid-liquid surface of W/P-GaN. As shown in Figure 3f, when N-GaN/ liquid/P-GaN junctions are excited under UV light, the majority carriers in N-GaN and photogenerated holes in P-GaN will diffuse to the solid-liquid interface. Simultaneously, the dynamic process of polarization/depolarization of liquid molecules continuously drive the circle of charged carriers in semiconductors as a result of dynamic screen effect.

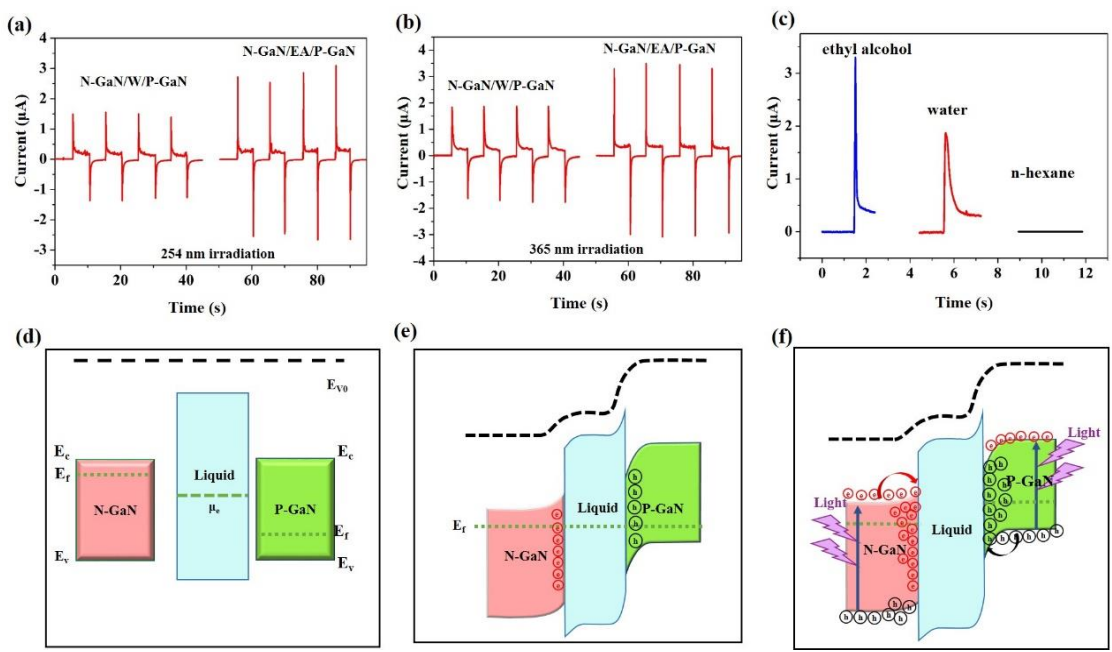

Figure 3. Time-resolved photo-polarization current of the N-GaN/W/P-GaN and N-GaN/EA/P-GaN photodetector with (a) 254 nm and (b) 365 nm switch on/off under zero voltage bias. (c) Output transient photo-polarization current of N-GaN/liquid/P-GaN photodetector under 365 nm illumination with different liquids, including ethylene glycol, water and nonpolar liquid n-hexane. The band structures schematic of the N-GaN/W/P-GaN photodetector (d) before contact, (e) during the contact state and (f) after light illumination.

The optical intensity dependent photo-polarization current characteristics were further explored by changing the light power density of the light source. As shown in Figure 4a, the polarized current is highly dependent on the optical power density. Due to the continuous polarization of liquid molecules under the increasement of the incident light power as more photo-induced charge carriers are accumulated at the interface between liquid and semiconductors, the response current increases gradually without a saturation, which means this type of photodetector could have a good performance competitive with the traditional PN diodes based photodetector.

The two indispensable parameters named as R and $D^*$, which represents the performance of the photodetector. In particular, the R and $D^*$ represents the ability to convert unit optical power into photo-current and detect weak signal. The R can be determined by the following formula (1):

$$R = \frac{I_L - I_D}{P} \qquad (1)$$

where $I_L$ represents the photo-polarization current, $I_D$ represents the dark current, $P$ represents the incident optical power. Meanwhile, $D^*$ can be determined by the following formula (2):

$$D^* = \frac{\sqrt{S} \times R}{\sqrt{2 \times q \times I_D}} \qquad (2)$$

where $S$ represents the active work area of detector, $q$ represents the charge constant. In the framework of solid-liquid photodetector, spike current and steady state current are two indispensable parameters. Figure. 4b shows the R and D$^*$ under steady state photo-polarization current for N-GaN/W/P-GaN photodetector as a function of incident light power density. The maximum R and D$^*$ are calculated as 7.4 mA/W and $2.2 \times 10^{11}$ Jones under zero bias voltage. Notably, photogenerated carriers might be trapped by defect states in the semiconductor to prolong the carrier lifetime under lower power densities whereas higher power densities bring decaying number of defect states causing relatively more exciton recombination[27, 28]. Meanwhile, R and D$^*$ of transient photo-polarization current for N-GaN/W/P-GaN photodetector was exhibited in Figure. 4c. The maximum R and D$^*$ are calculated as 104.2 mA/W and $4.4 \times 10^{12}$ Jones under zero bias voltage.

Breaking the limitations of requirement of lattice match in the material growth process, the photo-polarized currents on N-GaN/liquid/P-Si N-GaN/W/P-Si, N-GaN/W/N-Si and N-GaN/W/P-GaAs photodetectors were successfully recorded by I-T test under the 254 nm irradiation with zero bias voltage is shown in Figure 4(d) to Figure 4(f), revealing the stability and uniformity of the photo-polarized current. The characteristic curves exhibit sharp transient polarization and depolarization current peaks, confirming the mechanistic inference described above. The transient and steady state photo-polarization current basically reached 0.241 μA and 0.0156 μA in N-GaN/W/P-Si, 0.0514 μA and 0.0126 μA in N-GaN/W/N-Si, as well as 0.0467 μA and 7.35 nA in N-GaN/W/P-GaAs. The transient and steady state photo-polarization current curves corresponding to the device with polarized liquid is ethyl alcohol are exhibited in Figure. S4 (Supporting Information). The photo-polarized current in N-GaN/W/P-GaAs is smaller than the others, which might be caused by the rapid recombination of carriers in the doped

GaAs[29]. It is worth noting that the transient photo-polarization current of N-GaN/W/P-Si and N-GaN/W/N-Si photodetector under 254 nm illumination ranges from 0.216/0.245 μA and 0.0463/0.0548 μA, respectively. We speculate that the value of photo-polarization current is related to the Fermi level difference of the semiconductor on opposite sides of the polar liquid. The Fermi level of the P/N-Si was determined by the following formula (3) and (4)[30]

$$\rho_{n/p} = \frac{1}{q \times N_{D/A} \times \mu_{n/p}} \tag{3}$$

$$E_{F-n/p} = E_i \pm k_B \times T \times In\frac{N_{D/A} - N_{A/D}}{n_i} \tag{4}$$

where $\rho$ represents the resistivity of the P/N-Si. $N_{D/A}$ represents the carrier concentration of P/N-Si. $\mu_n$ (1450 cm$^2$/Vs) represents the intrinsic electron mobility of the N-Si and $\mu_p$ (500 cm$^2$/Vs) represents the intrinsic hole mobility of the P-Si. $E_i$ (4.61 eV) represents the median of the conduction and valence bands, $k_B$ represents the Boltzmann constant, $T$ represents the temperature and $n_i$ (1.5×10$^{10}$ cm$^{-3}$) represents the intrinsic electron concentration of the Si. Therefore, Fermi levels of the N-Si and P-Si were calculated as 4.32 eV and 4.93 eV, respectively. Meanwhile, the work function of N-GaN and P-GaN is around ~4.0 eV and ~6.1 eV. Based on I-T characteristic curves, we can verify that the larger Fermi level difference will drive more carriers to diffuse to the solid-liquid interface at the instant of illumination. In order to further intuitively exhibit the above inference, Figure S5 shows the relationship between the polar liquid and transient photo-polarization current in photodetectors with four sets of semiconductors. Obviously, when the polarized liquid is water, the transient photo-polarization current is smaller than that of ethyl alcohol in N-GaN/liquid/P-GaN, N-GaN/liquid/P-Si, N-GaN/liquid/N-Si and N-GaN/liquid/P-GaAs photodetector. According to these measurements of the transient photo-polarization current, we successfully detected electrical signals in response to ultraviolet light by inserting polar liquids into semiconductors.

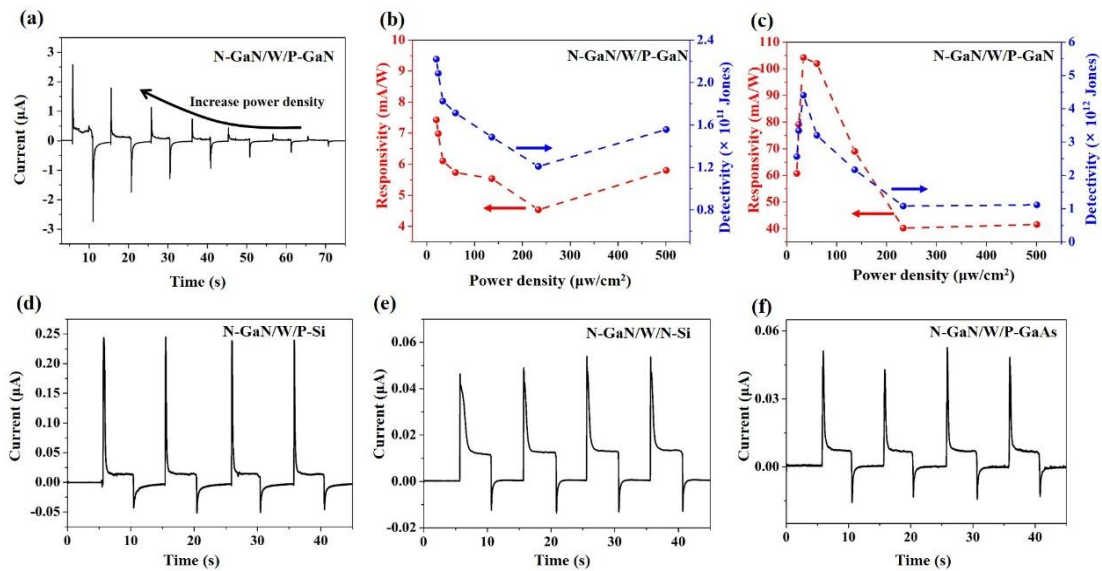

Figure 4. (a) The photo-polarization current of the N-GaN/W/P-GaN device measured under 365 nm illumination with different optical power intensities. Responsivity and detectivity of (b) steady photo-polarization current and (c) transient photo-polarization current as a function of incident optical power intensity at zero bias. The photo-polarization current of (d) N-GaN/W/P-Si, (d) N-GaN/W/N-Si, and (e) N-GaN/W/P-GaAs photodetector under 254 nm.

## Conclusion

In summary, an extremely convenient method to realize liquid-semiconductor mix-phase ultraviolet photodetector is presented by integrating polar liquid between the n-type and p-type semiconductors. Through inserting liquid into the PN junction, the photoexcited carriers will continuously diffuse to the opposite sides of the polar liquid droplet, and device will output stable photo-polarization current as an equilibrium is achieved between the polarization and depolarization process. Under 365 nm illumination, the responsivity and detectivity of transient photo-polarized current and steady state photo-polarized current in N-GaN/water/P-GaN reach values of 104.2 mA/W and $4.4 \times 10^{12}$ Jones, 7.4 mA/W and $2.2\times10^{11}$ Jones, respectively. This study provides the potential way to break through the constraints on the lattice match of heterojunction semiconductors photodetector, where we can flexibly choose the photodetector with response wavelength.

## Experimental Section

***Device fabrication and measurement***: Deionized water, containing a few foreign ions $\sigma < 20$ μS/cm, was obtained by lab water purification system (Heal Force RWD-1). Ethyl alcohol and n-hexane in this work were analytical grade without further purification. The P-GaN film with Mg concentration $\sim5\times10^{19}$ cm$^{-3}$ and thickness $\sim4$ μm was epitaxial grown on a sapphire substrate (Nanowin, China). The N-GaN film with Si concentration $\sim1.3\times10^{18}$ cm$^{-3}$ and thickness $\sim600$ nm was epitaxial grown on a sapphire substrate (HC SemiTek). In order to ensure Ohmic contacts between the electrode and GaN, Ni (15 nm)/Au (75 nm) contact was thermally evaporated onto one side of substrates successively. Similarly, Ti/Au (5 nm/70 nm) contact was thermally evaporated on the unpolished back side of GaAs substrate. The electrode of Si was grown using the magnetron sputtering metal gold on the unpolished back side, which formed an ohmic contact after annealing in Ar gas. Subsequently, a 150 μm thick square polyimide type with a 3 × 4 mm window was transferred onto pre-prepared semiconductors. All substrates need to be ultrasonically cleaned by acetone, isopropanol, ethyl alcohol and deionized water for 5 minutes before used.

***Characterization analysis:*** The I-V curves were measured by a Keithley 2400 source meter and I-T curves were carried out by Keithley 6514 source meter and DMM6500 multimeter. Raman spectra were measured by a Reinishaw system with an excitation laser of 532 nm. Power density of 254 nm and 365 nm light was detected by an S120 VC optical power meter by THOR-LABSU. UV-vis absorption spectra of GaN was recorded by UV-vis spectrophotometer (NatureSci Technologies Corporati Lambda950) in the range of 300-1000 nm. SEM images were obtained by ZEISS Utral 55.

## Acknowledgements

S.S.L. thanks the support from the National Natural Science Foundation of China (No. 51202216, 51502264, 61774135, and 51991342), Outstanding Youth Fund of Zhejiang Natural Science Foundation of China (LR21F040001), and Special Foundation of Young Professor of Zhejiang University (2013QNA5007). Y.H.L. thanks the support from the China Postdoctoral Science Foundation (2021M692767).